\documentclass[twocolumn,pra,amsfonts,aps,superscriptaddress]{revtex4-1}
\begin{document}
\title{Comment on `Experimental retrodiction of trajectories in double interferometer''}
\author{Marcin Wie\'sniak}\affiliation{Instutute of Informatics, Faculty of Mathematics, Physics and Informatics, University of Gda\'nsk,\\80-308 Gda\'nsk, Poland.}
\maketitle
In a recent article \cite{ERT} Yuan {\em et al.} perform another experiment themed on marking a path inside an interferometer. On page one, column right, they write:
\begin{quote}
Especially, two experiments obtained anomalous trajectories of photons not always following continuous trajectories \cite{VAIDMAN,ZHOU}  [References relabelled-M.W.]. The experiments were performed using an asymmetric Mach-Zehnder interferometer (MZI) with a symmetric MZI inserted into one arm, and used weak interactions to mark the path that photons take through the interferometer, where the experimenta	l results are explained in the framework of the two-state vector formalism of quantum theory \cite{AV1,AV2,AV3}. Afterwards the analysis of the experiment in \cite{VAIDMAN} using standard quantum optical methods and an amendment version were proposed \cite{POTOCEK,ENGLERT}. 
\end{quote}

By ``not always following continuous trajectories'' the Authors refer to a claim made by the Authors of Ref. \cite{VAIDMAN}. The article described an experiment with nested Mach-Zehnder interferometer, with a smaller loop inserted into the upper arm of the larger loop. The corners of the smaller loop and the lower arm of the large one are contributed with mirrors A, B, C, which vibrate with frequencies $f_A, f_B, f_C$, whereas mirrors E and F help to couple the light to and from the smaller loop and vibrate at $f_E$ and $f_F$. In a particular configuration of such an interferometer, Danan, Farfurnik, Bar-Ad, and Vaidman, saw peaks at the former three frequencies in the power spectrum of the signal reaching a detector at the end of the large loop, but not at the latter two.

However, as noticed by Sokolovsky \cite{SOKOL}, this lack does not imply that light in any way has omitted some of the elements of the setup. The absence of light at these locations is a no-go statement, which requires disproving {\em all} possible effects of the light being there. Recently, a detailed interferometric description \cite{SPECTRA} has pointed out features of the spectra related to frequencies of oscillations of all relevant elements of the setup.

One may, naturally, argue that the two-state vector formalism only describes first-order effects. This would imply that as an approximate theory it does not allow us to draw such definite conclusions. Another approach would include the fact that Vaidman (the main theorist of Ref. \cite{VAIDMAN}) introduced a gradation of presence \cite{VAIDMAN1} with {\em full} presence at a certain location being related to presence of certain peaks in a power spectrum (or nonvanishing of certain quantities \cite{LYING}), and {\em secondary} presence, unknown in any major physical theory, which accommodates all other consequences of interaction between light and an object. Even if we agree that the light was only secondarily present at mirrors in front of and behind the smaller Mach-Zehnder loop, this would not imply discontinuous trajectories. One could only say that photons were not present at these mirrors {\em only in Vaidman's sense}, but customizing such fundamental notions only contributes to the confusion.

This work has been supported by a NCN grant. NO 2015/19/B/ST2/0199.

\end{document}